\documentclass[a4paper,reqno]{amsart}

\usepackage{mathrsfs,url}
\let\mathcal\mathscr

\usepackage[matrix,arrow,tips,ps,dvips]{xy}

\SelectTips{cm}{}

\allowdisplaybreaks[1]

\newcommand*{\pd}[2]{\mathchoice{\frac{\partial#1}{\partial#2}}
  {\partial#1/\partial#2}{\partial#1/\partial#2}
  {\partial#1/\partial#2}}
\newcommand*{\fd}[2]{\mathchoice{\frac{\delta#1}{\delta#2}} {\delta
    #1/\delta#2}{\delta#1/\delta#2}{\delta#1/\delta#2}}

\newcommand{\enVert}[2][\right]{\relax \ifx#1\right\relax
\left\lVert\else#1\lVert\fi#2#1\rVert} \let\matr=\enVert
 
\let\phi\varphi

\newcommand{\cprime}{\/{\mathsurround=0pt$'$}}

\DeclareFontFamily{OML}{cyi}{} \DeclareFontShape{OML}{cyi}{m}{n}{ <5>
  <6> <7> <8> <9> gen * wncyi <10> <10.95> <12> <14.4> <17.28> <20.74>
  <24.88> wncyi10 }{} \DeclareSymbolFont{rusletters}{OML}{cyi}{m}{n}
\DeclareSymbolFontAlphabet{\rusmath}{rusletters}
\DeclareMathSymbol\re{\rusmath}{rusletters}{"03}

\DeclareMathOperator{\gf}{\mathbf{gf}}
\DeclareMathOperator{\sym}{\mathbf{sym}}

\providecommand{\href}[2]{#2} \providecommand{\urlprefix}{URL }
\providecommand*{\eprint}[2][]{%
\href{http://arXiv.org/abs/#2}{\begingroup \Url{arXiv:#2}}%
}

\theoremstyle{remark}
\newtheorem{remark}{Remark}

\begin{document}

\title[The $N=1$ supersymmetric KdV equation]{(Non)local Hamiltonian
  and symplectic structures, recursions, and hierarchies: a new
  approach and applications to the $N=1$ supersymmetric KdV equation}

\author[P. Kersten]{P. Kersten*}

\address{Paul Kersten \\
  University of Twente, Faculty of Electrical Engineering, Mathematics
  and Computer Science \\
  P.O.~Box 217 \\
  7500 AE Enschede \\
  The Netherlands}

\email{kersten@math.utwente.nl}

\thanks{*Corresponding author}

\author{I. Krasil{\cprime}shchik}

\address{Iosif Krasil{\cprime}shchik \\
  The Diffiety Institute and Independent University of Moscow \\
  B. Vlasevsky~11 \\
  119002 Moscow \\
  Russia}

\email{josephk@diffiety.ac.ru}

\author{A.~Verbovetsky}

\address{Alexander Verbovetsky \\
  Independent University of Moscow \\
  B. Vlasevsky~11 \\
  119002 Moscow \\
  Russia}

\email{verbovet@mccme.ru}



\begin{abstract}
  Using methods of \cite{KerstenKrasilshchikVerbovetsky:HOpC} and
  \cite{KrasilshchikKersten:SROpCSDE}, we accomplish an extensive
  study of the $N = 1$ supersymmetric Korteweg-de Vries equation. The
  results include: a description of local and nonlocal Hamiltonian and
  symplectic structures, five hierarchies of symmetries, the
  corresponding hierarchies of conservation laws, recursion operators
  for symmetries and generating functions of conservation laws. We
  stress that the main point of the paper is not just the results on
  super-KdV equation itself, but merely exposition of the efficiency
  of the geometrical approach and of the computational algorithms
  based on it.
\end{abstract}
\maketitle

\section*{Introduction}
There exists a number of super extensions of the classical KdV
equation
\begin{equation*}
  u_t=-u_{xxx}+6uu_{x}
\end{equation*}
(see \cite{Mathieu:OpPSKEq} and the references therein). One of them,
the so-called $N=1$ supersymmetric extension, is
\begin{align}\label{eq:1}
  u_t&=-u_{xxx}+6uu_{x}+\phi_{xx}\phi,\nonumber\\
  \phi_t&=-\phi_{xxx}+3u\phi_{x}+3u_{x}\phi,
\end{align}
where $\phi$ is an odd (fermionic) variable,~\cite{Oevel-Pop:1991}. To
deal with this system, it is convenient to introduce a new independent
odd variable $\theta$ such that $D_\theta^2=D_x$, where
\begin{equation*}
  D_\theta=\partial_\theta+\theta D_x
\end{equation*}
(here $D_x$ denotes the total derivative operator; see below) and a
new odd function
\begin{equation*}
  \Phi=\phi+\theta u.
\end{equation*}
Then \eqref{eq:1} will acquire the form
\begin{equation}\label{eq:2}
  \Phi_t=-\Phi_{xxx}+3D_\theta(\Phi)\Phi_{x}+3D_\theta(\Phi_x)\Phi.
\end{equation}
This equation is linear in $\theta$ and reduces to~\eqref{eq:1} if we
equal to each other the corresponding coefficients at the left- and
right-hand sides. System~\eqref{eq:1} (or equation~\eqref{eq:2}) was
studied before (see, e.g., \cite{ManinRadul:SExKPH}) and a number of
results related to its integrability were obtained. The aim of our paper
is twofold: (1)~to represent the known results in a more convenient form
(at least, from our point of view); (2)~to demonstrate the efficiency of
new methods of analysis of integrable systems described
in~\cite{KerstenKrasilshchikVerbovetsky:HOpC,
  KrasilshchikKersten:SROpCSDE} and based on a general geometric
approach to nonlinear PDE~\cite{KrasilshchikVinogradov:SCLDEqMP,
  KrasilshchikVerbovetsky:HMEqMP}. Actually, description of \emph{these
  methods and their highly algorithmical nature} (and, to a less extent,
of the results on the super-KdV equation themselves) is the main goal of
the paper. For traditional approach to Hamiltonian formalism in
integrable systems we refer the reader
to~\cite{Dorfman:1993,Magri:1978,Vin:1978}; an extensive exposition of
the theory for superintegrable systems can be found
in~\cite{Kuper:1987}.

This paper is organized as follows. In Section~\ref{section1}, we
present the essential definitions and results needed for applications
paying main attention to the computational aspects rather than to
theoretical ones. All the proofs can be found
in~\cite{KrasilshchikVinogradov:SCLDEqMP,
  KerstenKrasilshchikVerbovetsky:HOpC, KrasilshchikKersten:SROpCSDE,
  KrasilshchikVerbovetsky:HMEqMP}. In Section~\ref{section2}, the
results for the $N=1$ supersymmetric KdV equation are described.
Finally, in the last section we briefly discuss the results and
perspectives.

\section{Description of the computational scheme}
\label{section1}

Here we deal with evolution systems $\mathcal{E}$ of the form
\begin{equation}\label{eq:3}
  v_t=F(y,t,v_1,\dots,v_k),
\end{equation}
where both the unknown variable $v=(v^1,\dots,v^m)$ and the right-hand
side $F = (F^1,\dots,F^m)$ are vector-functions and $v_i =\pd{^i
  v}{y^i}$, $y$ and $t$ being the independent variables.
\begin{remark}\label{remark}
  In applications, some of the variables $v^j$, as well as $y$, may be
  \emph{odd}.  In particular, in equation \eqref{eq:2} $\theta$ and
  $\Phi$ are odd and $x$ is even.  Nevertheless, for the sake of
  simplicity, we expose the general theory for purely \emph{even}
  equations. Necessary corrections needed for the super case the reader
  will find in Subsection~\ref{subsection1.10}.
\end{remark}
Two basic operators related to \eqref{eq:3},
\begin{equation*}
  D_y=\pd{}{y}+\sum_{i,j} v^j_{i+1} \pd{}{v^j_i},\quad
  D_t=\pd{}{t}+\sum_{i,j} D^i_{y}(F^j)\pd{}{v^j_i},
\end{equation*}
are called the \emph{total derivatives}.

\begin{remark}\label{rem:functions}
  Note that the above expressions for total derivatives contain infinite
  number of terms. To make the action of these operators (as well as of
  similar operators introduced below) well defined, we introduce the
  space $\mathcal{F}(\mathcal{E})$ of functions smoothly depending on
  $y$, $t$ and a \emph{finite number} of variables $v^j_i$, and assume
  $D_y$ and $D_t$ to act in this space. Similarly, we shall consider the
  spaces $\mathcal{F}^m(\mathcal{E})$ of vector-functions of length $m$
  that depend on $y$, $t$ and $v^j_i$ in the same way.
\end{remark}

\subsection{Symmetries}
\label{subsection1.1}

A \emph{symmetry} of equation \eqref{eq:3} is a vector field
\begin{equation*}
  S=\sum_{i,j} S^j_i\pd{}{v^j_i},\quad S_i^j\in\mathcal{F}(\mathcal{E}),
\end{equation*}
such that
\begin{equation*}
  [S, D_y] = [S, D_t] = 0.
\end{equation*}
Any symmetry is of the form
\begin{equation}\label{eq:re_f}
  \re_f=\sum_{i,j}D^i_y(f^j)\pd{}{v^j_i},
\end{equation}
where the vector-function $f =
(f^1,\dots,f^m)\in\mathcal{F}^m(\mathcal{E})$ satisfies the system of
equations
\begin{equation}\label{eq:4}
  D_t(f^l)=\sum_{i,j}\pd{F^l}{v^j_i}D^i_y(f^j),\quad   l=1,\dots,m.
\end{equation}
The operator at the right-hand side of \eqref{eq:4} is called the
\emph{linearization} of $F$ and is denoted by $\ell_F$. Thus, equation
\eqref{eq:4} acquires the form
\begin{equation}\label{eq:5}
  D_t(f)=\ell_F(f).
\end{equation}

There exists a one-to-one correspondence between
symmetries~\eqref{eq:re_f} and the corresponding functions
$f\in\mathcal{F}^m(\mathcal{E})$, hence we shall identify symmetries
with such functions and use the term `symmetry' for any function that
satisfy~\eqref{eq:5}.

\subsection{Conservation laws and generating functions}
\label{subsection1.2}

A \emph{conservation law} of system \eqref{eq:3} is a pair
$\Omega=(Y,T)$, $Y$, $T\in\mathcal{F}(\mathcal{E})$, such that
\begin{equation}\label{eq:6}
  D_t(Y) = D_y(T).
\end{equation}
The function $Y$ is called the \emph{density} of $\Omega$. A
conservation law is called \emph{trivial} if $Y = D_y(P)$, $T =
D_t(P)$ for some function $P\in \mathcal{F} (\mathcal{E})$.

To any conservation law there corresponds its \emph{generating
  function} defined by
\begin{equation*}
  g_{\Omega}=\fd{Y}{v}=\biggl(\fd{Y}{v^1},\dots,\fd{Y}{v^m}\biggr),
\end{equation*}
where
\begin{equation*}
\fd{}{v^j}=\sum_{i \ge 0}(-D_y)^i\circ \pd{}{v^j_i}
\end{equation*}
is the \emph{variational derivative} with respect to $v^j$. Generating
functions of conservation laws satisfy the system of equations
\begin{equation}\label{eq:8}
  D_t(g)=-\ell_F^{*}(g),
\end{equation}
or
\begin{equation}\label{eq:7}
  D_t(g^l)=- \sum_{i,j}(-D_y)^i\biggl(\pd{F^j}{v^l_i}g^j\biggr),
  \quad   l=1,\dots,m,
\end{equation}
where $\ell_F^{*}$ is \emph{adjoint} to the operator $\ell_F$.

Any conservation law is uniquely determined by its generating function
and, in particular, $\Omega$ is trivial if and only if $g_{\Omega}=0$.
Stress that equation~\eqref{eq:7} may possess solutions that do not
correspond to any conservation law of~\eqref{eq:3}.

\begin{remark}
  Generating functions are also called
  \emph{cosymmetries}~\cite{Blaszak:1998} or \emph{conserved
    covariants}~\cite{FuchFok:1981}.
\end{remark}

\subsection{Nonlocal variables}
\label{subsection1.3}

Let us introduce a set of variables $w^1,\dots,w^j,\dots$ satisfying
the equations
\begin{equation}\label{eq:9}
  w^j_y=A^j(y,t,\dots,v^{\alpha}_i,\dots,w^{\beta},\dots),
  \quad w^j_t=B^j(y,t,\dots,v^{\alpha}_i,\dots,w^{\beta},\dots),
\end{equation}
that are compatible modulo equation \eqref{eq:3}, where $A^j$, $B^j$
are some smooth functions depending on a finite number of arguments.
Consider the operators
\begin{equation*}
  \widetilde{D}_y=D_y+\sum_{j}A^j\pd{}{w^j},\quad
  \widetilde{D}_t=D_t+\sum_{j}B^j\pd{}{w^j}.
\end{equation*}
Due to the compatibility conditions, one has
\begin{equation}
  \label{eq:PhysA-N1-v2:1}
  [\widetilde{D}_y,\widetilde{D}_t]=0
\end{equation}
modulo~\eqref{eq:3}. The variables $w^j$ are called \emph{nonlocal}.

Using the operators $\widetilde{D}_y$, $\widetilde{D}_t$ instead of
$D_y$ and $D_t$ in formulas \eqref{eq:4}, \eqref{eq:6}, and
\eqref{eq:7}, we can introduce the notions of \emph{nonlocal
  symmetries}, \emph{nonlocal conservation laws}, and \emph{nonlocal
  generating functions} depending on the new variables $w^j$. We shall
denote the spaces of such symmetries and generating functions by
$\sym(\mathcal{E})$ and $\gf(\mathcal{E})$, respectively.

\begin{remark}\label{rem:coverings}
  An invariant geometric way to introduce nonlocal variables is based
  on the notion of \emph{covering}, see~\cite{
    KrasilshchikVinogradov:SCLDEqMP,
    KrasilshchikKersten:SROpCSDE,KrasilshchikVinogradov:NTGDEqSCLBT,
    KrasilshchikVerbovetsky:HMEqMP}.
\end{remark}

\subsection{The $\ell$- and $\ell^*$-extensions}
\label{subsection1.4}

There are two canonical ways to extend the initial system
\eqref{eq:3}. The first one is related to the operator $\ell_F$ and is
called the \emph{$\ell$-extension}. Namely, let us introduce the
nonlocal variables $\omega^j_i$ (we shall also denote $\omega^j_0$ by
$\omega^j$), $j=1,\dots,m$, $i=0$, $1$, $\dots$\,, satisfying the
relations
\begin{equation*}
  (\omega^j_i)_y=\omega^j_{i+1}, \quad
  (\omega^j_i)_t=\widetilde{D}_y^i\biggl(\sum_{s,l}
  \pd{F^j}{v^l_s}\omega^l_s\biggr).
\end{equation*}
Clearly, these equations are consistent modulo \eqref{eq:3} and are
the consequences of the following ones
\begin{equation}\label{eq:10}
  \omega^j_t=\sum_{i,l} \frac{\partial F^j}{\partial v^l_i}\omega^l_i.
\end{equation}

In a similar way we construct the \emph{$\ell^{*}$-extension}: the
nonlocal variables are $p^j_i$ ($p^j_0$ will also be denoted by $p^j$)
and the defining relations are
\begin{equation*}
  (p^j_i)_y=p^j_{i+1},\quad
  (p^j_i)_t=-\widetilde{D}_y^i\biggl(\sum_{s,l}(-\widetilde{D}_y)^s
  \biggl(\pd{F^l}{v^j_s}p^l\biggr)\biggr),
\end{equation*}
that reduce to the equations
\begin{equation}\label{eq:11}
  p^j_t=-\sum_{s,l}(-\widetilde{D}_y)^s\biggl(\pd{F^l}{v^j_s}p^l\biggr)
\end{equation}
and their differential consequences.

\begin{remark}\label{rem:parities}
  The parities of the variables $\omega^j$ and $p^j$ are opposite to
  that of $v^j$: if $v^j$ is \emph{even}, then $\omega^j$ and $p^j$
  are \emph{odd} and vice versa.
\end{remark}

If the initial equation $\mathcal{E}$ was extended by nonlocal
variables $w^j$, we can associate to these variables, in a canonical
way, the corresponding $\omega$'s and $p$'s whose `behavior' is
governed by linearization or, respectively, adjoint linearization of
equations~\eqref{eq:9} in the corresponding nonlocal setting.

\subsubsection*{Associating operators to functions on the $\ell$- and
  $\ell^{*}$-extensions}

Let $\mathcal{F}^m(\mathcal{E})$ be the space of vector-valued
functions of length $m$ (see Remark~\ref{rem:functions}). Consider the
case when $\mathcal{E}$ is not extended by nonlocal variables first.
Let $a = (a_1,\dots,a_m)$, $a_i = \sum_{jl} a^{ij}_l\omega^j_l$,
$a^{ij}_l\in\mathcal{F}(\mathcal{E})$, be a linear in $\omega$
vector-function. Then we put into correspondence to this function a
differential operator $\Delta_a
=\Vert\Delta_a^{ij}\Vert\colon\mathcal{F}^m (\mathcal{E})\to
\mathcal{F}^m (\mathcal{E})$, where
\begin{equation*}
  \Delta^{ij}_a=\sum_l a^{ij}_l D^l_y,\qquad i,j=1,\dots,m.
\end{equation*}

If $\mathcal{F}(\mathcal{E})$ contains nonlocal variables, the
situation becomes more complicated. We shall consider here the
simplest case when the functions $A^j$ in \eqref{eq:9} are independent
of $\omega^{\beta}$. Let $\bar{\omega}^\beta$ be the variable in the
$\ell$-extension associated to the nonlocal variable $w^{\beta}$ and
$b =(b^1,\dots, b^m)$, $b^i = \sum_{\beta}
b^{i\beta}\bar{\omega}^\beta$, be a linear in $\bar{\omega}$
vector-function.  Then the corresponding operator
$\Delta_b=\Vert\Delta_b^{ij}\Vert\colon \mathcal{F}^m
(\mathcal{E})\to\mathcal{F}^m (\mathcal{E})$ is of the form
\begin{equation}\label{eq:12}
  \Delta^{ij}_b=\sum_{\alpha} b^{i\alpha}D_y^{-1}\circ\sum_{l}
  \pd{A^{\alpha}}{v^j_l} D_y^l.
\end{equation}
For the $\ell^{*}$-extension the construction is completely similar.

Below we shall use the notation $\mathcal{L}^m (\ell_{\mathcal{E}})$
and $\mathcal{L}^m(\ell^{*}_{\mathcal{E}})$ for the spaces of
vector-functions linear in $\omega$, $\bar{\omega}$ and $p$,
$\bar{p}$, respectively.

\subsection{Recursion operators for symmetries}
\label{subsection1.5}

Let $R\in\mathcal{L}^m(\ell_\mathcal{E})$ be a function that
satisfies the equation
\begin{equation*}
  \widetilde{D}_t(R)=\widetilde{\ell}_F(R).
\end{equation*}
Then the corresponding operator $\Delta_R$ maps
$\sym(\mathcal{E})$ to $\sym(\mathcal{E})$ and thus is a recursion
operator for (nonlocal) symmetries of $\mathcal{E}$.

\begin{remark}
  Here and below by $\widetilde{\ell}_F$ (or $\widetilde{\ell}_F$) we
  denote the linearization operator (or the adjoint one) in which the
  total derivatives~$D_y$ are substituted by the
  operators~$\widetilde{D}_y$ in the $\ell$- or $\ell^*$-extensions.
\end{remark}

\subsection{Recursion operators for generating functions}
\label{subsection1.6}

Let $L\in\mathcal{L}^m(\ell_\mathcal{E}^*)$ be a function that
satisfies the equation
\begin{equation*}
  \widetilde{D}_t(L) = -\widetilde{\ell}_F^{*}(L).
\end{equation*}
Then the corresponding operator $\Delta_L$ maps $\gf(\mathcal{E})$ to
$\gf(\mathcal{E})$ and thus is a recursion operator for (nonlocal)
generating functions of $\mathcal{E}$ (or \emph{adjoint recursion
  operator}~\cite{Blaszak:1998}).

\subsection{Hamiltonian structures}
\label{subsection1.7}

Let $K\in\mathcal{L}^m(\ell_{\mathcal{E}}^{*})$ be a function that
satisfies the equation
\begin{equation*}
  \widetilde{D}_t(K) = \widetilde{\ell}_F(K).
\end{equation*}
Then the corresponding operator $\Delta_K$ maps $\gf(\mathcal{E})$ to
$\sym(\mathcal{E})$. We call such maps \emph{pre-Hamiltonian
  structures} (they are also known as \emph{Noether
  operators}~\cite{FuchFok:1981}). In order $\Delta_K$ to be a true
\emph{Hamiltonian structure}, it has to satisfy two conditions:
skew-symmetry ($\Delta_K^{*}=-\Delta_K$) and the Jacobi identity for
the corresponding Poisson bracket (that amounts to
$[\![\Delta_K,\Delta_K]\!]=0$, where $[\![\cdot,\cdot]\!]$ is the
\emph{variational Schouten bracket}, see
\cite{IgoninVerbovetskyVitolo:FLVDOp,KerstenKrasilshchikVerbovetsky:HOpC}).
Both these conditions are easily checked in terms of the function~$K$.

Namely, if $K=\matr[\big]{\sum_{jl} a^{ij}_l p^j_l}$ then we consider
the function $W_K=\sum_{ijl}a_l^{ij}p^j_lp^i$ and in terms of $W_K$
the first condition reads
\begin{equation}\label{eq:13}
  \sum_{i}\fd{W_K}{p^i}p^i=-2W_K,
\end{equation}
while the second one is
\begin{equation}\label{eq:14}
  \biggl(\fd{}{v},\fd{}{p}\biggr)
  \sum_{i}\biggl(\fd{W_K}{v^i} \fd{W_K}{p^i}\biggr)=0,
\end{equation}
$(\fd{}{v},\fd{}{p}) =(\fd{}{v^1},\dots,\fd{}{v^m},
\fd{}{p^1},\dots,\fd{}{p^m})$. Note also that the compatibility
condition for two Hamiltonian structures $K$ and $K'$ amounts to
\begin{equation}\label{eq:15}
  \biggl(\fd{}{v},\fd{}{p}\biggr)\sum_{i}\biggl(\fd{W_K}{v^i} \fd{W_{K'}}{p^i}
  + \fd{W_{K'}}{v^i} \fd{W_K}{p^i}\biggr)=0.
\end{equation}

The equation~$\mathcal{E}$ itself is in the Hamiltonian form if it
possesses a Hamiltonian structure $K$ and may be presented as
\begin{equation}\label{eq:16}
  v_t=\Delta_K\fd{Y}{v}
\end{equation}
for some $Y=(Y^1,\dots,Y^m)$.

\subsection{Symplectic structures}
\label{subsection1.8}

Let $J\in \mathcal{L}^m(\ell_{\mathcal{E}})$ be a function that
satisfies the equation
\begin{equation*}
  \widetilde{D}_t(J) =-\widetilde{\ell}^{*}_F(J).
\end{equation*}
Then the corresponding operator $\Delta_J$ maps $\sym(\mathcal{E})$ to
$\gf(\mathcal{E})$ and may be called a \emph{presymplectic structure}
on $\mathcal{E}$ (alternatively, \emph{inverse Noether
  operator~\cite{FuchFok:1981}}). A presymplectic structure is called
\emph{symplectic} if it enjoys in addition the following properties.
Let $J=\matr[\big]{\sum_{jl}b^{ij}_l\omega^j_l}$.  Similar to
Subsection~\ref{subsection1.7}, we consider the function $W_J =
\sum_{ijl}b^{ij}_l\omega^j_l\omega^i$ and impose the conditions
\begin{equation}\label{eq:17}
  \sum_i \fd{W_J}{\omega^i}\omega^i=-2W_J,
\end{equation}
i.e., the operator $\Delta_J$ is skew-adjoint, and
\begin{equation}\label{eq:18}
  \biggl(\fd{}{v},\fd{}{\omega}\biggr)\sum_i\fd{W_J}{v^i} \omega^i=0
\end{equation}
that means that the `form' $W_J$ s is closed. Thus, in our context the
term `symplectic structure' means the same as in classical mechanics,
cf.~\cite{MarsdenRatiu:1999}.

\subsection{Canonical representation}
\label{subsection1.9}

As it will be seen below, all the operators constructed in our study
are presented in the form
\begin{equation*}
  \sum_{\alpha \ge 0} c^{\alpha}_{ij}D_{y}^{\alpha}
  +\sum_{\beta}d^{\beta}_{j} D_{y}^{-1}\circ e^{\beta}_{i},
\end{equation*}
where $\matr[\big]{c^{\alpha}_{ij}}$ is an $m\times m$-matrix,
$\matr[\big]{d^{\beta}_{j}}$ is an $m\times l$-matrix, and
$\matr[\big]{e^{\beta}_{i}}$ is an $l\times m$-matrix for some $l>0$
(matrix-valued functions, to be more precise).  In the table it is
shown how the matrices $d$ and $e$ look for different types of
operators.

\medskip
\begin{tabular}{|l|c|c|}
  \hline
  Type of operator                 & Lines of matrix $d$   &   Columns of matrix $e$ \\
  \hline
  Recursions for symmetries        &   Symmetry            &   Generating function   \\
  Recursions for generating funct. &   Generating function &   Symmetry        \\
  Hamiltonian structures           & Symmetry              &  Symmetry     \\
  Symplectic structures            &   Generating function & Generating function     \\
  \hline
\end{tabular}

\subsection{Super case}
\label{subsection1.10}

We shall now assume that all objects under consideration belong to the
super setting, i.e., may be either even or odd, which means that they
obey the rule
\begin{equation*}
  AB=(-1)^{AB}BA.
\end{equation*}
Here and below, symbols used at the exponents of $(-1)$ stand for the
corresponding~parity.  Generalization of the above exposed theory to
the super case is carried out along the lines
of~\cite{Verbovetsky:LFGAl,KrasilshchikVerbovetsky:HMEqMP}.

Then the basic formulas to be used in the calculus described above
are:
\begin{enumerate}
\item for evolutionary derivations
  \begin{equation*}
    \re_\phi=\sum_{ij}(-1)^{\phi v_i^j}D_y^i(\phi^j)\pd{}{v_i^j}
  \end{equation*}
  (naturally, the parity of $v_i^j$ equals that of $v^j$ plus parity
  of $y$ times $i$);
\item for the linearization one has
  $\ell_f(\phi)=(-1)^{f\phi}\re_\phi(f)$ that amounts to
  \begin{equation*}
    (\ell_f)_\alpha^\beta
    =\sum_i(-1)^{(f^\alpha+1)v_i^\beta}\pd{f^\alpha}{v_i^\beta}D_y^i;
  \end{equation*}
\item for the operator adjoint to $\Delta=\sum_ia_iD_y^i$ one has
  \begin{equation*}
    \Delta^*=\sum_i(-1)^{i+ia_iy+\frac{i(i-1)}{2}y}D_y^i\circ a_i.
  \end{equation*}
\end{enumerate}

\section{Main results for the $N = 1$ supersymmetric KdV equation}
\label{section2}

Here we apply the theory described above to equation~\eqref{eq:2}
\begin{equation*}
  \Phi_t=-\Phi_{xxx}+3D_\theta(\Phi)\Phi_x+3D_\theta(\Phi_x)\Phi.
\end{equation*}
We use the notation
\begin{align*}
  \Phi_{k} &\quad\text{ for }\quad\pd{^{2k}\Phi}{\theta^{2k}}
  =\pd{^{k}\Phi}{x^k} \intertext{and} \Phi_{k\frac{1}{2}} &\quad\text{
    for }\quad D_\theta^{2k+1}(\Phi)
  =D_\theta\bigg(\pd{^k\Phi}{x^k}\bigg).
\end{align*}
The functions $\Phi_{k}$ are \emph{odd} while $\Phi_{k\frac{1}{2}}$
are \emph{even}, the function $\Phi=\Phi_0$ itself being odd.

\subsubsection*{Gradings}

We assign the following gradings $[\cdot]$ to the variables on our
equation:
\begin{equation*}
  [\theta]=-{1}/{2},\ \ [x]=-1,\ \ [t]=-3,\ \ [\Phi]={3}/{2}.
\end{equation*}
Respectively, we have
\begin{equation*}
  [\Phi_{k}]=(2k+3)/{2},\ \ \ [\Phi_{k \frac{1}{2}}]=k+2.
\end{equation*}
With these gradings, equation \eqref{eq:2} becomes homogeneous (of
grading $9/2$) and all constructions below can be considered to be
homogeneous as well.

\subsection{Nonlocal functions}
\label{subsection2.1}

Here we extend the equation~$\mathcal{E}$ by four groups of nonlocal
variables. We present here their $\theta$-components only; the $x$-
and $t$-components are given
in~\cite{KerstenKrasilshchikVerbovetsky:ExSNSKEq} (they are found from
compatibility conditions~\eqref{eq:PhysA-N1-v2:1}).

\subsubsection{Group 1}
\label{subsection2.1.1}

This group includes the even variables $q_1$, $q_3$, $q_5$, defined by
\begin{align*}
  (q_1)_{\theta}&=\Phi_{0},\\
  (q_3)_{\theta}&=\Phi_{0} \Phi_{\frac{1}{2}},\\
  (q_5)_{\theta}&=\Phi_{\frac{1}{2}}
  (-\Phi_{2}+2\Phi_{0}\Phi_{\frac{1}{2}})/2.
\end{align*}

\noindent
\emph{Gradings}: $[q_1] = 1$, $[q_3] = 3$, $[q_5] = 5$.

\subsubsection{Group 2}
\label{subsection2.1.2}

This group includes the odd variables $Q_{\frac{1}{2}}$,
$Q_{\frac{5}{2}}$, $Q_{\frac{9}{2}}$ defined by
\begin{align*}
  (Q_{\frac{1}{2}})_{\theta}&=q_1,\\
  (Q_{\frac{5}{2}})_{\theta}&=q_1^3-6q_3,\\
  (Q_{\frac{9}{2}})_{\theta}&=-60 \Phi_0\Phi_1 q_1 + q_1^5 - 60 q_1^2
  q_3+240 q_5.
\end{align*}

\noindent
\emph{Gradings}: $[Q_{\frac{1}{2}}]={1}/{2}$,
$[Q_{\frac{5}{2}}]={5}/{2}$, $[Q_{\frac{9}{2}}]={9}/{2}$.

\subsubsection{Group 3}
\label{subsection2.1.3}

This group includes the odd variables $Q_{\frac{3}{2}}$,
$Q_{\frac{7}{2}}$, $Q_{\frac{11}{2}}$ defined by
\begin{align*}
  (Q_{\frac{3}{2}})_\theta&=\Phi_0 Q_{\frac{1}{2}},\\
  (Q_{\frac{7}{2}})_\theta&=(12 \Phi_2 Q_{\frac{1}{2}}+ 18
  \Phi_1Q_{\frac{1}{2}} q_1
  +\Phi_0 Q_{\frac{5}{2}})/3,\\
  (Q_{\frac{11}{2}})_\theta&
  =(360\Phi_4Q_{\frac12}+5280\Phi_3Q_{\frac12}q_1-760\Phi_2Q_{\frac52}
  +4680\Phi_2Q_{\frac12}\Phi_{\frac12}+1200\Phi_2Q_{\frac12}q_1^2\\
  &+60\Phi_1Q_{\frac52}q_1+\Phi_0Q_{\frac92})/60.
\end{align*}

\noindent
\emph{Gradings}: $[Q_{\frac{3}{2}}]={3}/{2}$,
$[Q_{\frac{7}{2}}]={7}/{2}$, $[Q_{\frac{11}{2}}]={11}/{2}$.

\subsubsection{Group 4}
\label{subsection2.1.4}

This group includes the even variables $\bar q_1$, $\bar q_3$, $\bar
q_5$ defined by
\begin{align*}
  ({\bar q_1})_{\theta}&=Q_{\frac{3}{2}},\\
  ({\bar q_3})_{\theta}&=-(Q_{\frac{7}{2}} + Q_{\frac{3}{2}} q_1^2),\\
  ({\bar q_5})_{\theta}&=( 12 Q_{\frac{11}{2}} + 42 Q_{\frac{7}{2}}
  \Phi_{\frac{1}{2}} + 6Q_{\frac{7}{2}} q_1^2 + 12 Q_{\frac{3}{2}}
  \Phi_{1 \frac{1}{2}} q_1 +Q_{\frac{3}{2}} q_1^4 -24 Q_{\frac{3}{2}}
  q_1 q_3)/3.
\end{align*}

\noindent
\emph{Gradings}: $[\bar q_1]=1$, $[\bar q_3]=3$, $[\bar q_5]=5$.

\begin{remark}\label{remark+}
  The last three variables are not used directly in the subsequent
  computations, but clarify the nonlocal picture and enter in the
  expressions for the higher terms of hierarchies of symmetries and
  generating functions.
\end{remark}

\subsection{Seed symmetries}
\label{subsection2.2}

Solving equation \eqref{eq:4}, which in our case is of the form
\begin{equation*}
  \widetilde{D}_t(f)=-\widetilde{D}_{\theta}^6(f) + 3
  \widetilde{D}_{\theta}(f)\Phi_1 + 3 \Phi_{\frac{1}{2}}
  \widetilde{D}_{\theta}^2(f) +3\widetilde{D}_{\theta}^3(f) \Phi+3
  \Phi_{1 \frac{1}{2}}f,
\end{equation*}
where $\widetilde{D}_\theta=\partial_\theta+\theta\widetilde{D}_x$,
while~$\widetilde{D}_x$ and~$\widetilde{D}_t$ are the total derivative
operators extended to the nonlocal setting (see Subsection
\ref{subsection2.1}), we found a number of solutions that serve as
seed symmetries for constructing infinite hierarchies and are used to
construct \emph{nonlocal vectors} (see Subsection \ref{subsection2.4}
below).

These symmetries are:

\subsubsection*{The $Y_k$ series}
\label{subsection2.2.1}

\begin{align*}
  Y_1&=\Phi_1,\\
  Y_3&=\Phi_3-3 \Phi_1 \Phi_{\frac{1}{2}} - 3 \Phi_0 \Phi_{1 \frac{1}{2}},\\
  Y_5&= \Phi_5-5 \Phi_3 \Phi_{\frac{1}{2}} -10 \Phi_2 \Phi_{1
    \frac{1}{2}} +10\Phi_1\Phi_{\frac12}^2 - 10 \Phi_1
  \Phi_{2\frac{1}{2}}
  + 20 \Phi_0 \Phi_{\frac{1}{2}} \Phi_{1 \frac{1}{2}} \\
  &- 5 \Phi_0\Phi_{3 \frac{1}{2}}.
\end{align*}

\subsubsection*{The $Y_{k\frac12}$ series}
\label{subsection2.2.2}

\begin{align*}
  Y_{\frac{3}{2}}&= - 2 \Phi_1 Q_{\frac{1}{2}} -\Phi_{\frac{1}{2}} q_1
  +  \Phi_{1\frac{1}{2}},\\
  Y_{\frac{7}{2}}&= -12 \Phi_3 Q_{\frac{1}{2}} - 2 \Phi_1
  Q_{\frac{5}{2}} + 36\Phi_1 Q_{\frac{1}{2}} \Phi_{\frac{1}{2}}
  + 36 \Phi_0 Q_{\frac{1}{2}} \Phi_{1\frac{1}{2}} +12 \Phi_0 \Phi_2 \\
  &- 6 \Phi_0 \Phi_1 q_1 + 12 \Phi_{\frac{1}{2}}^2q_1 - 36
  \Phi_{\frac{1}{2}} \Phi_{1\frac{1}{2}} - \Phi_{\frac{1}{2}} q_1^3 +
  6\Phi_{\frac{1}{2}} q_3 +3 \Phi_{1\frac{1}{2}} q_1^2
  -6 \Phi_{2\frac{1}{2}} q_1\\
  &+ 6 \Phi_{3\frac{1}{2}},\\
  Y_{\frac{11}{2}}&=
  240\Phi_5Q_{\frac12}+40\Phi_3Q_{\frac52}-1200\Phi_3Q_{\frac12}\Phi_{\frac12}
  -2400\Phi_2Q_{\frac12}\Phi_{1\frac12}\\
  &+2\Phi_1Q_{\frac92}
  -120\Phi_1Q_{\frac52}\Phi_{\frac12}+2400\Phi_1Q_{\frac12}\Phi_{\frac12}^2
  -2400\Phi_1Q_{\frac12}\Phi_{2\frac12}
  -600\Phi_1\Phi_3\\
  &+240\Phi_1\Phi_2q_1-120\Phi_0Q_{\frac52}\Phi_{1\frac12}
  +4800\Phi_0Q_{\frac12}\Phi_{\frac12}\Phi_{1\frac12}
  -1200\Phi_0Q_{\frac12}\Phi_{3\frac12}\\
  &-480\Phi_0\Phi_4+360\Phi_0\Phi_3q_1+1920\Phi_0\Phi_2\Phi_{\frac12}
  -120\Phi_0\Phi_2q_1^2
  -720\Phi_0\Phi_1\Phi_{\frac12}q_1\\
  &+1680\Phi_0\Phi_1\Phi_{1\frac12}+20\Phi_0\Phi_1q_1^3
  -120\Phi_0\Phi_1q_3+660\Phi_{\frac12}^3q_1
  -3540\Phi_{\frac12}^2\Phi_{1\frac12}\\
  &-40\Phi_{\frac12}^2q_1^3
  +240\Phi_{\frac12}^2q_3+360\Phi_{\frac12}\Phi_{1\frac12}q_1^2
  -960\Phi_{\frac12}\Phi_{2\frac12}q_1
  +1200\Phi_{\frac12}\Phi_{3\frac12}\\
  &+\Phi_{\frac12}q_1^5-60\Phi_{\frac12}q_1^2q_3
  +240\Phi_{\frac12}q_5-720\Phi_{1\frac12}^2q_1
  +2400\Phi_{1\frac12}\Phi_{2\frac12}
  -5\Phi_{1\frac12}q_1^4\\
  &+120\Phi_{1\frac12}q_1q_3+20\Phi_{2\frac12}q_1^3
  -120\Phi_{2\frac12}q_3-60\Phi_{3\frac12}q_1^2
  +120\Phi_{4\frac12}q_1-120\Phi_{5\frac12}.
\end{align*}

\subsubsection*{The $Z_k$ series}
\label{subsection2.2.3}

\begin{align*}
  Z_1&=Q_{\frac{1}{2}} \Phi_{\frac{1}{2}} +\theta( -2 \Phi_1
  Q_{\frac{1}{2}}
  -\Phi_{\frac{1}{2}} q_1 +\Phi_{1\frac{1}{2}} ),\\
  Z_3&=( 3 Q_{\frac{3}{2}} \Phi_{\frac{1}{2}} q_1- 3 Q_{\frac{3}{2}}
  \Phi_{1\frac{1}{2}} + Q_{\frac{5}{2}} \Phi_{\frac{1}{2}} - 12
  Q_{\frac{1}{2}} \Phi_{\frac{1}{2}}^2- 3 Q_{\frac{1}{2}}
  \Phi_{1\frac{1}{2}} q_1
  + 6 Q_{\frac{1}{2}} \Phi_{2\frac{1}{2}}\\
  &+ 6 \Phi_1 Q_{\frac{1}{2}} Q_{\frac{3}{2}}+ 6 \Phi_0 \Phi_1
  Q_{\frac{1}{2}} +\theta(-12 \Phi_3Q_{\frac{1}{2}}
  - 2 \Phi_1 Q_{\frac{5}{2}}+ 36 \Phi_1 Q_{\frac{1}{2}} \Phi_{\frac{1}{2}}\\
  &+ 36 \Phi_0 Q_{\frac{1}{2}} \Phi_{1\frac{1}{2}}+ 12 \Phi_0 \Phi_2-
  6 \Phi_0 \Phi_1 q_1
  + 12 \Phi_{\frac{1}{2}}^2 q_1- 36 \Phi_{\frac{1}{2}} \Phi_{1\frac{1}{2}}\\
  &- \Phi_{\frac{1}{2}} q_1^3+ 6 \Phi_{\frac{1}{2}} q_3+ 3
  \Phi_{1\frac{1}{2}}q_1^2
  -6\Phi_{2\frac{1}{2}}q_1 + 6 \Phi_{3\frac{1}{2}})/3,\\
  Z_5&= (- 15 Q_{\frac{7}{2}} \Phi_{\frac{1}{2}} q_1+15
  Q_{\frac{7}{2}} \Phi_{1\frac{1}{2}} + 120 Q_{\frac{3}{2}}
  \Phi_{\frac{1}{2}}^2 q_1
  -360 Q_{\frac{3}{2}} \Phi_{\frac{1}{2}} \Phi_{1\frac{1}{2}}\\
  &-10 Q_{\frac{3}{2}} \Phi_{\frac{1}{2}} q_1^3+ 60 Q_{\frac{3}{2}}
  \Phi_{\frac{1}{2}} q_3 + 30 Q_{\frac{3}{2}} \Phi_{1\frac{1}{2}}
  q_1^2- 60 Q_{\frac{3}{2}} \Phi_{2\frac{1}{2}}q_1
  + 60 Q_{\frac{3}{2}} \Phi_{3\frac{1}{2}}\\
  &- Q_{\frac{9}{2}} \Phi_{\frac{1}{2}}+ 40 Q_{\frac{5}{2}}
  \Phi_{\frac{1}{2}}^2 - 5 Q_{\frac{5}{2}} \Phi_{\frac{1}{2}} q_1^2
  +15 Q_{\frac{5}{2}} \Phi_{1\frac{1}{2}} q_1- 20 Q_{\frac{5}{2}}
  \Phi_{2\frac{1}{2}}
  - 660 Q_{\frac{1}{2}} \Phi_{\frac{1}{2}} ^3\\
  &+ 90 Q_{\frac{1}{2}} \Phi_{\frac{1}{2}}^2 q_1^2 - 390
  Q_{\frac{1}{2}} \Phi_{\frac{1}{2}} \Phi_{1\frac{1}{2}} q_1 +960
  Q_{\frac{1}{2}} \Phi_{\frac{1}{2}} \Phi_{2\frac{1}{2}}
  +  5 Q_{\frac{1}{2}} \Phi_{\frac{1}{2}}  q_1^4\\
  &- 30 Q_{\frac{1}{2}} \Phi_{\frac{1}{2}} q_1 q_3+660
  Q_{\frac{1}{2}}\Phi_{1\frac{1}{2}}^2 - 10 Q_{\frac{1}{2}}
  \Phi_{1\frac{1}{2}} q_1^3 -30Q_{\frac12}\Phi_{1\frac12}q_3
  + 60 Q_{\frac{1}{2}} \Phi_{3\frac{1}{2}} q_1\\
  & - 120 Q_{\frac{1}{2}} \Phi_{4\frac{1}{2}} +12 \Phi_5 -120 \Phi_3
  Q_{\frac{1}{2}} Q_{\frac{3}{2}}
  - 60 \Phi_3 \Phi_{\frac{1}{2}} -120 \Phi_2 \Phi_{1\frac{1}{2}}\\
  & - 20\Phi_1 Q_{\frac{5}{2}} Q_{\frac{3}{2}} - 30 \Phi_1
  Q_{\frac{1}{2}} Q_{\frac{7}{2}} + 360 \Phi_1 Q_{\frac{1}{2}}
  Q_{\frac{3}{2}} \Phi_{\frac{1}{2}}
  - 10 \Phi_1 Q_{\frac{1}{2}} Q_{\frac{5}{2}} q_1\\
  & - 240 \Phi_1 \Phi_2 Q_{\frac{1}{2}} - 60 \Phi_1 \Phi_{\frac{1}{2}}
  q_1^2 + 60 \Phi_1 \Phi_{1\frac{1}{2}} q_1 - 120 \Phi_1
  \Phi_{2\frac{1}{2}}
  +  360 \Phi_0 Q_{\frac{1}{2}} Q_{\frac{3}{2}} \Phi_{1\frac{1}{2}}\\
  & - 360 \Phi_0 \Phi_3 Q_{\frac{1}{2}} + 120 \Phi_0 \Phi_2
  Q_{\frac{3}{2}} + 120 \Phi_0 \Phi_2 Q_{\frac{1}{2}} q_1
  + 60\Phi_0\Phi_1 Q_{\frac32}q_1\\
  & - 20 \Phi_0 \Phi_1 Q_{\frac{5}{2}} + 720 \Phi_0 \Phi_1
  Q_{\frac{1}{2}} \Phi_{\frac{1}{2}} - 180 \Phi_0 \Phi_1
  Q_{\frac{1}{2}} q_1^2
  + 300 \Phi_0 \Phi_{\frac{1}{2}} \Phi_{1\frac{1}{2}}\\
  & - 90 \Phi_0 \Phi_{\frac{1}{2}} q_1^3 + 90 \Phi_0
  \Phi_{1\frac{1}{2}} q_1^2 -60\Phi_0\Phi_{3\frac12}
  + \theta(240\Phi_5Q_{\frac12}+40\Phi_3Q_{\frac52}\\
  &-1200\Phi_3Q_{\frac12}\Phi_{\frac12}
  -2400\Phi_2Q_{\frac12}\Phi_{1\frac12}+2\Phi_1Q_{\frac92}
  -120\Phi_1Q_{\frac52}\Phi_{\frac12}+2400\Phi_1Q_{\frac12}\Phi_{\frac12}^2\\
  &-2400\Phi_1Q_{\frac12}\Phi_{2\frac12}-600\Phi_1\Phi_3+240\Phi_1\Phi_2q_1
  -120\Phi_0Q_{\frac52}\Phi_{1\frac12}
  +4800\Phi_0Q_{\frac12}\Phi_{\frac12}\Phi_{1\frac12}\\
  &-1200\Phi_0Q_{\frac12}\Phi_{3\frac12}-480\Phi_0\Phi_4+360\Phi_0\Phi_3q_1
  +1920\Phi_0\Phi_2\Phi_{\frac12}-120\Phi_0\Phi_2q_1^2\\
  &-720\Phi_0\Phi_1\Phi_{\frac12}q_1
  +1680\Phi_0\Phi_1\Phi_{1\frac12}+20\Phi_0\Phi_1q_1^3
  -120\Phi_0\Phi_1q_3+660\Phi_{\frac12}^3q_1\\
  &-3540\Phi_{\frac12}^2\Phi_{1\frac12}
  -40\Phi_{\frac12}^2q_1^3+240\Phi_{\frac12}^2q_3
  +360\Phi_{\frac12}\Phi_{1\frac12}q_1^2
  -960\Phi_{\frac12}\Phi_{2\frac12}q_1\\
  &+1200\Phi_{\frac12}\Phi_{3\frac12}
  +\Phi_{\frac12}q_1^5-60\Phi_{\frac12}q_1^2q_3
  +240\Phi_{\frac12}q_5-720\Phi_{1\frac12}^2q_1
  +2400\Phi_{1\frac12}\Phi_{2\frac12}\\
  &-5\Phi_{1\frac12}q_1^4+120\Phi_{1\frac12}q_1q_3+20\Phi_{2\frac12}q_1^3
  -120\Phi_{2\frac12}q_3-60\Phi_{3\frac12}q_1^2
  +120\Phi_{4\frac12}q_1\\
  &-120\Phi_{5\frac12}))/5.
\end{align*}

\subsubsection*{The $Z_{k\frac12}$ series}
\label{subsection2.2.4}

\begin{align*}
  Z_{\frac{1}{2}}&=- 2 \theta \Phi_1+ \Phi_{\frac{1}{2}},\\
  Z_{\frac{5}{2}}&=- 2 \Phi_1 Q_{\frac{3}{2}}+ \Phi_1 Q_{\frac{1}{2}}
  q_1+ 2 \Phi_0 \Phi_1 - 4 \Phi_{\frac{1}{2}}^2+ \Phi_{\frac{1}{2}}
  q_1^2- 2 \Phi_{1\frac{1}{2}} q_1
  + 2 \Phi_{2\frac{1}{2}}\\
  &+ \theta(- 4 \Phi_3 + 12 \Phi_1\Phi_{\frac{1}{2}}
  + 12 \Phi_0 \Phi_{1\frac{1}{2}}),\\
  Z_{\frac{9}{2}}&=- 24 \Phi_3 Q_{\frac{3}{2}}+ 24 \Phi_3
  Q_{\frac{1}{2}} q_1 - 6 \Phi_1 Q_{\frac{7}{2}} + 72 \Phi_1
  Q_{\frac{3}{2}} \Phi_{\frac{1}{2}}
  + 2 \Phi_1 Q_{\frac{5}{2}} q_1\\
  &- 36 \Phi_1 Q_{\frac{1}{2}} \Phi_{\frac{1}{2}} q_1+ 24 \Phi_1
  Q_{\frac{1}{2}} \Phi_{1\frac{1}{2}} -36\Phi_1 Q_{\frac12}q_3
  + 48 \Phi_1 \Phi_2+ 72 \Phi_0 Q_{\frac{3}{2}} \Phi_{1\frac{1}{2}}\\
  &- 72 \Phi_0 Q_{\frac{1}{2}} \Phi_{1\frac{1}{2}} q_1+ 72 \Phi_0
  \Phi_3- 48 \Phi_0 \Phi_2 q_1
  - 144 \Phi_0 \Phi_1  \Phi_{\frac{1}{2}}+ 48 \Phi_0 \Phi_1 q_1^2\\
  &+ \theta(- 48 \Phi_5+ 240 \Phi_3 \Phi_{\frac{1}{2}}+ 480 \Phi_2
  \Phi_{1\frac{1}{2}} - 480 \Phi_1 \Phi_{\frac{1}{2}}^2
  + 480 \Phi_1 \Phi_{2\frac{1}{2}}\\
  &- 960 \Phi_0 \Phi_{\frac{1}{2}} \Phi_{1\frac{1}{2}}+ 240 \Phi_0
  \Phi_{3\frac{1}{2}}) + 132 \Phi_{\frac{1}{2}}^3 - 24
  \Phi_{\frac{1}{2}}^2 q_1^2
  + 144 \Phi_{\frac{1}{2}} \Phi_{1\frac{1}{2}} q_1\\
  &- 192 \Phi_{\frac{1}{2}} \Phi_{2\frac{1}{2}}+ \Phi_{\frac{1}{2}}
  q_1^4 - 24 \Phi_{\frac{1}{2}} q_1 q_3 - 144 \Phi_{1\frac{1}{2}} ^2-
  4 \Phi_{1\frac{1}{2}} q_1^3
  + 24 \Phi_{1\frac{1}{2}} q_3\\
  &+12 \Phi_{2\frac{1}{2}} q_1^2- 24 \Phi_{3\frac{1}{2}} q_1+ 24
  \Phi_{4\frac{1}{2}}.
\end{align*}

\subsubsection*{Gradings}

There are two points of view on symmetries: as on functions and as on
vector fields $\re_f$ (see Subsection~\ref{subsection1.1}).  For
functions we have:
\begin{align*}
  [Y_1]   &={5}/{2},&[Y_3]   &={9}/{2},  &[Y_5]    &={13}/{2}, &&\text{odd};\\
  [Y_{\frac{3}{2}}]&=3, &[Y_{\frac{7}{2}}]&=5, &[Y_{\frac{11}{2}}]&=7,
  &&\text{even};\\
  [Z_1]   &={5}/{2},&[Z_3]   &={7}/{2},  &[Z_5]    &= {13}/{2}, &&\text{odd};\\
  [Z_{\frac{1}{2}}]&=2, &[Z_{\frac{5}{2}}]&=4, &[Z_{\frac{9}{2}}] &=6,
  &&\text{even}.
\end{align*}

For vector fields we have:
\begin{align*}
  [\re_{Y_1}]&=1, &[\re_{Y_3}]&=3, &[\re_{Y_5}]&=5,  &&\text{even};\\
  [\re_{Y_{\frac{3}{2}}}]&= {3}/{2}, &[\re_{Y_{\frac{7}{2}}}
  ]&={7}/{2},
  &[\re_{Y_{\frac{11}{2}}}]&= {11}/{2},  &&\text{odd};\\
  [\re_{Z_{1}}]&= 1, &[\re_{Z_{3}}]&= 3, &[\re_{Z_{5}}]&= 5,  &&\text{even};\\
  [\re_{Z_{\frac{1}{2}}}]&= {1}/{2}, &[\re_{Z_{\frac{5}{2}}} ]&=
  {5}/{2}, &[\re_{Z_{\frac{9}{2}}}]&= {9}/{2}, &&\text{odd}.
\end{align*}

Note also that the symmetries $Y_{\alpha}$ do not depend on $\theta$,
while $Z_{\alpha}$ are linear functions with respect to $\theta$.

\subsection{Seed generating functions}
\label{subsection2.3}

Solving equation~\eqref{eq:4}, which in our case is of the form
\begin{equation*}
  \widetilde{D}_t(f) = - \widetilde{D}_{\theta}^6(f) +
  6\Phi_{\frac12}\widetilde{D}_\theta^2(f) - 3\Phi_0
  \widetilde{D}_\theta^3(f),
\end{equation*}
we found a number of solutions that serve as seed generating functions
for constructing infinite hierarchies and used to construct
\emph{nonlocal forms} (see Subsection~\ref{subsection2.5} below).
These generating functions are:

\subsubsection*{The $F_k$ series}
\label{subsection2.3.1}

\begin{align*}
  F_0 &= 1,\\
  F_2 &=  \Phi_{\frac{1}{2}},\\
  F_4&=(- 2\Phi_{0} \Phi_{1}+3 \Phi_{\frac{1}{2}}^2 -
  \Phi_{2\frac{1}{2}})/3.\hphantom{\hspace{6.5cm}}
\end{align*}

\subsubsection*{The $F_{k\frac12}$ series}
\label{subsection2.3.2}

\begin{align*}
  F_{\frac{1}{2}} &= Q_{\frac{1}{2}},\\
  F_{\frac{5}{2}} &= (Q_{\frac{5}{2}} - 12
  Q_{\frac{1}{2}}\Phi_{\frac{1}{2}} +6 \Phi_{1}
  +6 \Phi_{0}q_1)/6,\\
  F_{\frac{9}{2}} &=(Q_{\frac{9}{2}}- 40 Q_{\frac{5}{2}}
  \Phi_{\frac{1}{2}} +720 Q_{\frac{1}{2}} \Phi_{\frac{1}{2}}^2
  -240 Q_{\frac{1}{2}}\Phi_{2\frac{1}{2}}+ 120 \Phi_{3} + 120 \Phi_{2}q_1\\
  &- 480 \Phi_{1} \Phi_{\frac{1}{2}} + 60 \Phi_{1} q_1^2 -480 \Phi_{0}
  \Phi_{1} Q_{\frac{1}{2}} -420 \Phi_0 \Phi_{\frac{1}{2}} q_1
  - 240 \Phi_{0} \Phi_{1\frac{1}{2}} +20 \Phi_0 q_1^3\\
  & -120 \Phi_{0} q_3)/20.
\end{align*}

\subsubsection*{The $G_k$ series}
\label{subsection2.3.3}

\begin{align*}
  G_0&=\theta Q_{\frac12},\\
  G_2&=(3Q_{\frac12}Q_{\frac32}+6\Phi_0Q_{\frac12}+\theta Q_{\frac52}
  -12\theta Q_{\frac12}\Phi_{\frac12}
  +6\theta\Phi_1+6\theta\Phi_0q_1)/3,\\
  G_4&=(-10Q_{\frac52}Q_{\frac32}+15Q_{\frac12}Q_{\frac72}
  +120Q_{\frac12}Q_{\frac32}\Phi_{\frac12}
  -5Q_{\frac12}Q_{\frac52}q_1-120\Phi_2Q_{\frac12}\\
  &-60\Phi_1Q_{\frac32}-60\Phi_0Q_{\frac32}q_1-20\Phi_0Q_{\frac52}
  +420\Phi_0Q_{\frac12}\Phi_{\frac12}+90\Phi_0Q_{\frac12}q_1^2\\
  &-120\Phi_0\Phi_1-\theta Q_{\frac92}+40\theta
  Q_{\frac52}\Phi_{\frac12} -720\theta Q_{\frac12}\Phi_{\frac12}^2
  +240\theta Q_{\frac12}\Phi_{2\frac12}-120\theta\Phi_3\\
  &-120\theta\Phi_{2}q_1+480\theta\Phi_{1}\Phi_{\frac12}-60\theta\Phi_{1}q_1^2
  +480\theta\Phi_{0}\Phi_{1}Q_{\frac12}+420\theta\Phi_{0}\Phi_{\frac12}q_1\\
  &+240\theta\Phi_{0}\Phi_{1\frac12}-20\theta\Phi_{0}q_1^3
  +120\theta\Phi_{0}q_3)/90.
\end{align*}

\subsubsection*{The $G_{k\frac12}$ series}
\label{subsection2.3.4}

\begin{align*}
  G_{-\frac12}&=\theta,\\
  G_{\frac32}&=-Q_{\frac32}+Q_{\frac12}q_1+2\Phi_{0}-4\theta\Phi_{\frac12},\\
  G_{\frac72}&=(3Q_{\frac72}-24Q_{\frac32}\Phi_{\frac12}-Q_{\frac52}q_1
  +6Q_{\frac12}\Phi_{\frac12}q_1
  -12Q_{\frac12}\Phi_{1\frac12} + 18Q_{\frac12}q_3-24\Phi_{2}\\
  &-12\Phi_{1}q_1+84\Phi_{0}\Phi_{\frac12}+6\Phi_{0}q_1^2
  +96\theta\Phi_{0}\Phi_{1}-144\theta\Phi_{\frac12}^2
  +48\theta\Phi_{2\frac12})/6.
\end{align*}

\subsubsection*{Gradings}

These generating functions have the following gradings and parities:
\begin{align*}
  [F_0] &= 0, &[F_2] &= 2, &[F_4] &= 4, &&\text{even};\\
  [F_{\frac12}] &= 1/2, &[F_{\frac52}] &= 5/2, &[F_{\frac92}]
  &= 9/2, &&\text{odd};\\
  [G_0] &= 0, &[G_2] &= 2, &[G_4] &= 4, &&\text{even};\\
  [G_{-\frac12}] &= -1/2, &[G_{\frac32}] &= 3/2, &[G_{\frac72}] &=
  7/2, &&\text{odd}.
\end{align*}

Note again that the generating functions $F_\alpha$ do not depend on
$\theta$, while $G_\alpha$ are linear functions with respect
to~$\theta$.

\subsection{Nonlocal vectors}
\label{subsection2.4}

We consider now to the $\ell^*$-extension of equation~\eqref{eq:2}. The
additional coordinates on this extension are denoted by $P = P_0$,
$P_{\frac12}$, $P_1$,~etc.

Now we introduce nonlocal variables in the $\ell^*$-extension that we
call \emph{nonlocal vectors} and which are defined by
\begin{align*}
  (P_{Y_1})_\theta&=Y_1P_0,&(P_{Y_3})_\theta&=Y_3P_0,&(P_{Y_5})_\theta
  &=Y_5P_0;\\
  (P_{Y_{\frac32}})_\theta&=Y_{\frac32}P_0,&(P_{Y_{\frac72}})_\theta
  &=Y_{\frac72}P_0,
  &(P_{Y_{\frac{11}2}})_\theta&=Y_{\frac{11}2}P_0;\\
  (P_{Z_1})_\theta&=Z_1P_0,&(P_{Z_3})_\theta&=Z_3P_0,&(P_{Z_5})_\theta
  &=Z_5P_0;\\
  (P_{Z_{\frac12}})_\theta&=Z_{\frac12}P_0,&(P_{Z_{\frac52}})_\theta
  &=Z_{\frac52}P_0, &(P_{Z_{\frac92}})_\theta&=Z_{\frac92}P_0,
\end{align*}
where the symmetries $Y_\alpha$ and $Z_\alpha$ were described in
Subsection~\ref{subsection2.2}.

The $x$- and $t$-components of these variables are given
in~\cite{KerstenKrasilshchikVerbovetsky:ExSNSKEq}.

\subsubsection*{Gradings}
The variable $P_0$ is even and we assign grading $0$ to it. Then $P_k$
are also even variables with $[P_k] = k$ while $P_{k\frac12}$ are odd
and $[P_{k\frac12}] = (2k+1)/2$. Consequently,
\begin{align*}
  [P_{Y_1}]&=2,&[P_{Y_3}]&=4,&[P_{Y_5}]&=6,&&\text{even};\\
  [P_{Y_{\frac32}}]&=5/2,&[P_{Y_{\frac72}}]&=9/2,&[P_{Y_{\frac{11}2}}]
  &=13/2,&&\text{odd};\\
  [P_{Z_1}]&=2,&[P_{Z_3}]&=4,&[P_{Z_5}]&=6,&&\text{even};\\
  [P_{Z_{\frac12}}]&=3/2,&[P_{Z_{\frac52}}]&=7/2,&[P_{Z_{\frac92}}]
  &=11/2,&&\text{odd}.
\end{align*}

\subsection{Nonlocal forms}
\label{subsection2.5}

Passing to the $\ell$-extension of equation \eqref{eq:2}, we introduce
the additional coordinates on this extension that are denoted by
$\Omega = \Omega_0$, $\Omega_{\frac12}$, $\Omega_1$,~etc.

Now we introduce nonlocal variables in the $\ell$-extension called
\emph{nonlocal forms} and described by
\begin{align*}
  (\Omega_{F_0})_\theta&=\Omega_0F_0,&(\Omega_{F_2})_\theta&
  =\Omega_0F_2,&(\Omega_{F_4})_\theta&=\Omega_0F_4;\\
  (\Omega_{F_{\frac12}})_\theta&
  =\Omega_0F_{\frac12},&(\Omega_{F_{\frac52}})_\theta&=\Omega_0F_{\frac52},
  &(\Omega_{F_{\frac{9}2}})_\theta&=\Omega_0F_{\frac{9}2};\\
  (\Omega_{G_0})_\theta&=\Omega_0G_0,&(\Omega_{G_2})_\theta&
  =\Omega_0G_2,&(\Omega_{G_4})_\theta&=\Omega_0G_4;\\
  (\Omega_{G_{-\frac12}})_\theta&
  =\Omega_0G_{-\frac12},&(\Omega_{G_{\frac32}})_\theta&=\Omega_0G_{\frac32},
  &(\Omega_{G_{\frac72}})_\theta&=\Omega_0G_{\frac72},
\end{align*}
where the generating functions $F_\alpha$ and $G_\alpha$ were
described in Subsection~\ref{subsection2.3}.

The $x$- and $t$-components of these variables are given in
\cite{KerstenKrasilshchikVerbovetsky:ExSNSKEq}.

\subsubsection*{Gradings}
The variable $\Omega_0$ is even and we assign grading $0$ to it. Then
$\Omega_k$ are also even variables with $[\Omega_k] = k$, while
$\Omega_{k\frac12}$ are odd and $[\Omega_{k\frac12}] = (2k+1)/2$.
Consequently,
\begin{align*}
  [\Omega_{F_0}]&=-1/2,&[\Omega_{F_2}]&=3/2,&[\Omega_{F_4}]&=7/2,
  &&\text{odd};\\
  [\Omega_{F_{\frac12}}]&
  =0,&[\Omega_{F_{\frac52}}]&=2,&[\Omega_{F_{\frac{9}2}}]&=4,&&\text{even};\\
  [\Omega_{G_0}]&=-1/2,&[\Omega_{G_2}]&=3/2,&[\Omega_{G_4}]&=7/2,
  &&\text{odd};\\
  [\Omega_{G_{-\frac12}}]&=-1,&[\Omega_{G_{\frac32}}]&
  =1,&[\Omega_{G_{\frac72}}]&=3,&&\text{even}.
\end{align*}

\subsection{Recursion operators for symmetries}
\label{subsection2.6}

Using the method described in Subsection \ref{subsection1.5}, we found
two nontrivial solutions of the linearized equation in the
$\ell$-extension enriched with nonlocal variables. The first one is
\begin{align*}
  R_1&=-Q_{\frac12}\Omega_{F_0}\Phi_{\frac12}
  -2\Phi_1\Omega_{G_0}-\Phi_1\Omega_{F_0}
  +2\Phi_1Q_{\frac12}\Omega_{G_{-\frac12}}\\
  &-2\Phi_0\Omega_{\frac12}+\theta\Omega_{F_0}\Phi_{\frac12}q_1
  -\theta\Omega_{F_0}\Phi_{1\frac12}
  +2\theta\Phi_1Q_{\frac12}\Omega_{F_0}\\
  &+2\theta\Phi_1\Omega_{F_{\frac12}}-\Omega_{F_{\frac12}}\Phi_{\frac12}
  +\Omega_{G_{-\frac12}}\Phi_{\frac12}q_1
  -\Omega_{G_{-\frac12}}\Phi_{1\frac12}\\
  &-2\Omega_0\Phi_{\frac12}+\Omega_2.
\end{align*}

The operator corresponding to the first solution is
\begin{align*}
  \Delta_{R_1}&=D_\theta^4-2\Phi_0 D_\theta-2\Phi_{\frac12}\\
  &-(Y_1+Z_1)D_\theta^{-1}\circ F_0-Z_{\frac12}D_\theta^{-1}\circ
  F_{\frac12} -Y_{\frac32}D_\theta^{-1}\circ G_{-\frac12}
  -2Y_{1}D_\theta^{-1}\circ G_0.
\end{align*}
The second solution is given
in~\cite{KerstenKrasilshchikVerbovetsky:ExSNSKEq} and corresponds to
the operator~$\Delta_{R_1}^2$.

\subsubsection*{Gradings} 
The operator $R_1$ is even and its grading is $2$.

\subsection{Recursion operators for generating functions}
\label{subsection2.7}

Using the method described in Subsection \ref{subsection1.6}, we found
three nontrivial solutions of the adjoint linearized equation in the
$\ell^*$-extension enriched with nonlocal variables. The first one~is
\begin{equation*}
  L_1=Q_{\frac12}P_{Z_{\frac12}}+2\Phi_0P_{\frac12}+\theta
  P_{Y_{\frac32}} +2\theta Q_{\frac12}P_{Y_1}
  -4\Phi_{\frac12}P_0+P_{Y_1}+P_{Z_1}+P_2.
\end{equation*}
The operator corresponding to the first solution is
\begin{align*}
  \Delta_{L_1}&=D_\theta^4+2\Phi_0 D_\theta-4\Phi_{\frac12}\\
  &+(F_0+2G_0)D_\theta^{-1}\circ Y_1+G_{-\frac12}D_\theta^{-1}\circ
  Y_{\frac32} +F_0 D_\theta^{-1}\circ
  Z_1+F_{\frac12}D_\theta^{-1}\circ Z_{\frac12}.
\end{align*}
The second and third solutions are given
in~\cite{KerstenKrasilshchikVerbovetsky:ExSNSKEq} and correspond to
the operators~$\Delta_{L_1}^2$ and~$\Delta_{L_1}^3$, resp.

\subsubsection*{Gradings}
The operator $L_1$ is even and its grading is $2$.

\subsection{Hamiltonian structures}
\label{subsection2.8}

Using the method described in Subsection \ref{subsection1.7}, we found
three nontrivial solutions of the linearized equation in the
$\ell^*$-extension enriched with nonlocal variables. The first one is
\begin{equation*}
  K_1=P_{2\frac12}-P_{\frac12}\Phi_{\frac12}-2\Phi_1P_0-3\Phi_0P_1.
\end{equation*}
The operator corresponding to the first solution is
\begin{equation*}
  \Delta_{K_1}=D_\theta^5-3\Phi_0D_\theta^2-\Phi_{\frac12}D_\theta
  -2\Phi_1.
\end{equation*}
This operator satisfies criteria \eqref{eq:13} and \eqref{eq:14} and
thus is Hamiltonian.  Moreover, there exists a conservation law
(corresponding to the nonlocal variable $q_3$)
\begin{align*}
  X&=\Phi_0\Phi_{\frac12},\\
  T&=-2\Phi_1\Phi_2+\Phi_0\Phi_3-9\Phi_0\Phi_1\Phi_{\frac12}
  +4\Phi_{\frac12}^3 -2\Phi_{\frac12}\Phi_{2\frac12}+\Phi_{1\frac12}^2
\end{align*}
such that our equation can be represented as
\begin{equation*}
  \Phi_t=\Delta_{K_1}\fd{}{\Phi}\Big(-\frac{1}{2}X\Big),
\end{equation*}
and so \eqref{eq:16} is also satisfied.

The second Hamiltonian structure is of the form
\begin{align*}
  K_2&=-P_{Z_{\frac12}}\Phi_{\frac12}q_1+P_{Z_{\frac12}}\Phi_{1\frac12}
  -P_{Y_{\frac32}}\Phi_{\frac12}
  +P_{4\frac12}-3P_{2\frac12}\Phi_{\frac12}-3P_{1\frac12}\Phi_{1\frac12}\\
  &+3P_{\frac12}\Phi_{\frac12}^2
  -P_{\frac12}\Phi_{2\frac12}-2Q_{\frac12}\Phi_{\frac12}P_{Y_1}
  -2\Phi_3P_0-7\Phi_2P_1-2\Phi_1Q_{\frac12}P_{Z_{\frac12}}\\
  &+9\Phi_1\Phi_{\frac12}P_0-2\Phi_1P_{Z_1}-9\Phi_1P_2-\Phi_0\Phi_1P_{\frac12}
  +13\Phi_0\Phi_{\frac12}P_1+7\Phi_0\Phi_{1\frac12}P_0\\
  &-5\Phi_0P_3+2\theta\Phi_1P_{Y_{\frac32}}+4\theta\Phi_1Q_{\frac12}P_{Y_1}
  +2\theta\Phi_{\frac12}q_1P_{Y_1}-2\theta\Phi_{1\frac12}P_{Y_1}.
\end{align*}
The corresponding operator is
\begin{align*}
  \Delta_{K_2}&=D_\theta^9-5\Phi_0D_\theta^6
  -3\Phi_{\frac12}D_\theta^5-9\Phi_1D_\theta^4
  -3\Phi_{1\frac12}D_\theta^3+(13\Phi_0\Phi_{\frac12}-7\Phi_2)D_\theta^2\\
  &+(3\Phi_{\frac12}^2-\Phi_{2\frac12}
  -\Phi_0\Phi_1)D_\theta+(9\Phi_1\Phi_{\frac12}
  +7\Phi_0\Phi_{1\frac12}-2\Phi_3)\\
  &+Y_{\frac32}D_\theta^{-1}\circ
  Z_{\frac12}-Z_{\frac12}D_\theta^{-1}\circ Y_{\frac32}
  -2Y_1D_\theta^{-1}\circ Z_1 -2Z_1D_\theta^{-1}\circ Y_1.
\end{align*}
The third solution is given
in~\cite{KerstenKrasilshchikVerbovetsky:ExSNSKEq} (see also
Remark~\ref{rem:interrelations} below).

\subsubsection*{Gradings}
The operator $\Delta_{K_1}$ is odd and of grading $5/2$. The operator
$\Delta_{K_2}$ is also odd and of grading $9/2$.

\subsection{Symplectic structures}
\label{subsection2.9}

Using the method described in Subsection \ref{subsection1.8}, we found
three nontrivial solutions of the adjoint linearized equation in the
$\ell$-extension enriched with nonlocal variables. The first one is
\begin{equation*}
  J_1=\Omega_{G_0}+\Omega_{F_0}-Q_{\frac12}\Omega_{G_{-\frac12}}+\theta
  Q_{\frac12}\Omega_{F_0} +\theta\Omega_{F_{\frac12}}.
\end{equation*}
The operator corresponding to the first solution is
\begin{equation*}
  \Delta_{J_1}=(F_0+G_0)D_\theta^{-1}\circ
  F_0+G_{-\frac12}D_\theta^{-1}\circ F_{\frac12}
  -F_{\frac12}D_\theta^{-1}\circ G_{-\frac12}+F_0 D_\theta^{-1}\circ
  G_0.
\end{equation*}
This operator satisfies criteria \eqref{eq:17} and \eqref{eq:18} and
thus is symplectic.

The second solution is of the form
\begin{align*}
  J_2&=(3\Omega_{G_2}-12\Omega_{G_0}\Phi_{\frac12}
  -12\Omega_{F_2}-12\Omega_{F_0}\Phi_{\frac12}
  +6\Omega_{1\frac12}-3Q_{\frac32}\Omega_{F_{\frac12}}\\
  &-Q_{\frac52}\Omega_{G_{-\frac12}}
  +3Q_{\frac12}Q_{\frac32}\Omega_{F_0}+3Q_{\frac12}\Omega_{F_{\frac12}}q_1
  +12Q_{\frac12}\Omega_{G_{-\frac12}}\Phi_{\frac12}\\
  &-3Q_{\frac12}\Omega_{G_{\frac32}}
  -6\Phi_1\Omega_{G_{-\frac12}}+6\Phi_0Q_{\frac12}\Omega_{F_0}
  +6\Phi_0\Omega_{F_{\frac12}}-6\Phi_0\Omega_{G_{-\frac12}}q_1\\
  &+6\Phi_0\Omega_{0}+\theta Q_{\frac52}\Omega_{F_0}-12\theta
  Q_{\frac12}\Omega_{F_2}
  -12\theta Q_{\frac12}\Omega_{F_0}\Phi_{\frac12}+6\theta\Phi_1\Omega_{F_0}\\
  &+6\theta\Phi_0\Omega_{F_0}q_1
  -12\theta\Omega_{F_{\frac12}}\Phi_{\frac12}+6\theta\Omega_{F_{\frac52}})/6.
\end{align*}
The corresponding operator is
\begin{align*}
  \Delta_{J_2}&=D_\theta^3+\Phi_0+(\frac{1}{2}G_2-
  2F_2)D_\theta^{-1}\circ F_0\\
  &-2(F_0+G_0)D_\theta^{-1}\circ F_2+\frac12
  G_{\frac32}D_\theta^{-1}\circ F_{\frac12}
  +G_{-\frac12}D_\theta^{-1}\circ F_{\frac52}\\
  &-2F_2D_\theta^{-1}\circ G_0+\frac12 F_0D_\theta^{-1}\circ G_2
  -6F_{\frac52}D_\theta^{-1}\circ G_{-\frac12} -\frac12
  F_{\frac12}D_\theta^{-1}\circ G_{\frac32}.
\end{align*}
The third solution is given
in~\cite{KerstenKrasilshchikVerbovetsky:ExSNSKEq} (see
Remark~\ref{rem:interrelations}).

\subsubsection*{Gradings}
The operator $\Delta_{J_1}$ is odd and of grading $-1/2$. The second
operator is also odd and its grading equals $3/2$.

\subsection{Interrelations}
\label{subsection2.10}

Using the symmetries computed in Subsection~\ref{subsection2.2} and
applying the recursion operator obtained in
Subsection~\ref{subsection2.6}, we get four infinite series of
(generally, nonlocal) symmetries
\begin{align*}
  &Y_{2k-1},&[Y_{2k-1}]&=(4k+1)/2,&&\text{odd},\\
  &Y_{\frac{4k-1}{2}},&[Y_{\frac{4k-1}{2}}]&=2k+1,&&\text{even},\\
  &Z_{2k-1},&[Z_{2k-1}]&=(4k+1)/2,&&\text{odd},\\
  &Z_{\frac{4k-3}{2}},&[Z_{\frac{4k-3}{2}}]&=2k,&&\text{even},
\end{align*}
$k=1$, $2$, $\dots$

In a similar war, using the results of Subsections~\ref{subsection2.3}
and~\ref{subsection2.7}, we get four infinite series of generating
functions
\begin{align*}
  &F_{2k-2},&[F_{2k-2}]&=2k-2,&&\text{even},\\
  &F_{\frac{4k-3}{2}},&[F_{\frac{4k-3}{2}}]&=(4k-3)/2,&&\text{odd},\\
  &G_{2k},&[G_{2k}]&=2k,&&\text{even},\\
  &G_{\frac{4k-5}{2}},&[G_{\frac{4k-5}{2}}]&=(4k-5)/2,&&\text{odd},
\end{align*}
$k=1$, $2$, $\dots$

These series are related to each other (up to rational coefficients)
by the operators of
Subsections~\ref{subsection2.6}--\ref{subsection2.9} in the following
way:

\begin{gather*}
  \xymatrix{ & Y_{2k-1} \ar[r]^{\Delta_{R_1}}
    \ar[d]^{\Delta_{J_1}} &
    Y_{2k+1} \ar[d]^{\Delta_{J_1}} \\
    F_{2k-2} \ar[r]_{\Delta_{L_1}} \ar[ru]^{\Delta_{K_1}} & F_{2k}
    \ar[r]_{\Delta_{L_1}} \ar[ru]_{\Delta_{K_1}} & F_{2k+2} } \qquad
  \xymatrix{ & Z_{2k-1} \ar[r]^{\Delta_{R_1}}
    \ar[d]^{\Delta_{J_1}} &
    Z_{2k+1} \ar[d]^{\Delta_{J_1}} \\
    G_{2k} \ar[r]_{\Delta_{L_1}} \ar[ru]^{\Delta_{K_1}} & G_{2k+2}
    \ar[r]_{\Delta_{L_1}} \ar[ru]_{\Delta_{K_1}} & G_{2k+4}
  } \\[1ex]
  \xymatrix{ & Y_{\frac{4k-1}{2}} \ar[r]^{\Delta_{R_1}}
    \ar[d]^{\Delta_{J_1}} &
    Y_{\frac{4k+3}{2}} \ar[d]^{\Delta_{J_1}} \\
    F_{\frac{4k-3}{2}} \ar[r]_{\Delta_{L_1}} \ar[ru]^{\Delta_{K_1}} &
    F_{\frac{4k+1}{2}} \ar[r]_{\Delta_{L_1}} \ar[ru]_{\Delta_{K_1}} &
    F_{\frac{4k+5}{2}} } \qquad
  \xymatrix{ & Z_{\frac{4k-3}{2}}
    \ar[r]^{\Delta_{R_1}} \ar[d]^{\Delta_{J_1}} &
    Z_{\frac{4k+1}{2}} \ar[d]^{\Delta_{J_1}} \\
    G_{\frac{4k-5}{2}} \ar[r]_{\Delta_{L_1}} \ar[ru]^{\Delta_{K_1}} &
    G_{\frac{4k-1}{2}} \ar[r]_{\Delta_{L_1}} \ar[ru]_{\Delta_{K_1}} &
    G_{\frac{4k+3}{2}} }
\end{gather*}

\begin{remark}
  \label{remark2}
  Actually, there exists another hierarchy of symmetries $S_{2k}$,
  $k=0$, $1$,~\dots\,, with the seed element
  \begin{equation*}
    S_0=6(-\Phi_3+3\Phi_1\Phi_{\frac12}+3\Phi\Phi_{1\frac12})t
    +2\Phi_1x+\theta\Phi_{\frac12}+3\Phi
  \end{equation*}
  (the scaling symmetry). All these symmetries are odd, linear with
  respect to $x$, $t$, and~$\theta$, and have grading
  $[S_{2k}]=(4k+3)/2$, see~\cite{Oevel:1987}.
\end{remark}

\begin{remark}[cf.~\cite{Blaszak:1998}]
  \label{rem:interrelations}
  Let us clarify the relations between the structures described above.
  First, it should be noted that the Hamiltonian structures~$K_1$
  and~$K_2$ are \emph{compatible}, i.e., their Schouten bracket
  vanishes (or, their linear combination is a Hamiltonian structure
  again). More over, they are related to each other by the rerecursion
  operator~$R_1$:
  $\Delta_{K_2}=\Delta_{R_1}\circ\Delta_{K_1}$. Consequently, an
  infinite series of (nonlocal) compatible Hamiltonian
  structures~$K_i$ arises, such that
  $\Delta_{K_{i+1}}=\Delta_{R_1}\circ\Delta_{K_i}$. In a similar way,
  we have an infinite series of symplectic structures related by the
  operator~$L_1$. The inverse of each Hamiltonian structure, if it
  makes sense, is a symplectic structure and vice versa.

  Second, in an obvious way all natural powers of recursion operators
  are also recursion operators. If~$R$ is a recursion operator for
  symmetries, the adjoint~$R^*$ is a recursion for generating
  functions and vice versa.
\end{remark}

\section{Conclusion}
\label{section3}

The study of the $N = 1$ supersymmetric KdV equation exposed in this
paper demonstrates the power and efficiency of the geometrical methods
elaborated in~\cite{KrasilshchikVinogradov:SCLDEqMP}
and~\cite{KerstenKrasilshchikVerbovetsky:HOpC}.  In particular, we
found recursion operators for symmetries and generating functions,
Hamiltonian and symplectic structures, constructed five infinite
series of symmetries. The research was based on new geometrical
methods giving rise to efficient computational algorithms.

Our experience shows that the methods applied are of a universal
nature and may be used to analyze a lot of other equations, both
classical and supersymmetric. In particular, from technical point of
view, the canonical representation of nonlocal operators (see
Subsection~\ref{subsection1.9}) seems to be quite efficient and
convenient when dealing with such operators. Note that all nonlocal
operators constructed in this paper are represented in the canonical
form.

We strongly believe that the majority of the problems formulated in
\cite{Mathieu:OpPSKEq} can be solved by our methods. We plan to
demonstrate this in forthcoming publications. Note in particular that
the nonlocal Hamiltonian structure indicated in \cite{Mathieu:OpPSKEq}
is inverse to our symplectic structure $J_1$.

\section*{Acknowledgments}

I.K. and A.V. are grateful to the University of Twente, where this
research was done, for hospitality. The work of A.V. was supported in
part by the NWO and FOM (The Netherlands). We also want to thank our
anonymous referee for his/her extremely valuable comments.

\end{document}